\documentclass{ws-mpla}
\usepackage{graphicx}
\usepackage{amsmath,amssymb}

\begin{document}

\markboth{V. Borka Jovanovi\'{c} et al.} {Masses of constituent
quarks confined in open bottom hadrons}

\catchline{}{}{}{}{}

\title{MASSES OF CONSTITUENT QUARKS CONFINED IN OPEN BOTTOM HADRONS}

\author{V. BORKA JOVANOVI\'{C}$^1$\footnote{vborka@vinca.rs}}

\author{D. BORKA$^1$}

\author{P. JOVANOVI\'{C}$^2$}

\author{J. MILO\v{S}EVI\'{C}$^{3,4}$}

\author{S. R. IGNJATOVI\'{C}$^5$}

\address{$^1$Atomic Physics Laboratory (040), Vin\v{c}a Institute of Nuclear Sciences, \\
University of Belgrade. P.O. Box 522, 11001 Belgrade, Serbia}

\address{$^2$Astronomical Observatory, Volgina 7, 11060 Belgrade, Serbia}

\address{$^3$Laboratory of Physics (010), Vin\v{c}a Institute of Nuclear Sciences, \\
University of Belgrade, P.O. Box 522, 11001 Belgrade, Serbia}

\address{$^4$Department of Physics, University of Oslo, Oslo, Norway}

\address{$^5$Department of Physics, Faculty of Science, University of Banja Luka, \\
Mladena Stojanovi\'{c}a 2, 78000 Banja Luka, Bosnia and Herzegovina}

\maketitle

\pub{Received (Day Month Year)}{Revised (Day Month Year)}

\begin{abstract}
We apply color-spin and flavor-spin quark-quark interactions to the
meson and baryon constituent quarks, and calculate constituent quark
masses, as well as the coupling constants of these interactions. The
main goal of this paper was to determine constituent quark
masses from light and open bottom hadron masses, using the fitting
method we have developed and clustering of hadron groups. We use
color-spin Fermi-Breit (FB) and flavor-spin Glozman-Riska (GR)
hyperfine interaction (HFI) to determine constituent quark masses
(especially $b$ quark mass). Another aim was to discern between the 
FB and GR HFI because our previous findings had indicated that both 
interactions were satisfactory. Our improved fitting procedure of
constituent quark masses showed that on average color-spin (Fermi-Breit)
hyperfine interaction yields better fits. The method also shows the
way how the constituent quark masses and the strength of the
interaction constants appear in different hadron environments.


\end{abstract}

\ccode{PACS Nos.: 12.39.Pn; 12.40.Yx; 14.20.Mr; 14.40.Nd}

\section{Introduction}
\label{sec1}

Determination of quark masses is extremely important, for both
phenomenological and theoretical applications.\cite{lubi01} About
the importance of the mass of the bottom quark, as a fundamental
parameter of the Standard Model, see also review by El-Khadra and
Luke.\cite{elkh02}

Many spectroscopic quark models, based on two-body interactions,
have been developed (see e.g. Ref.~\refcite{gonz08} and references
therein). These are called ``hyperfine interaction'' (HFI) in
analogy with the atomic physics. The simplest HFI is the spin-spin
interaction proposed by Rujula, Georgi and Glashow back in 
1975,\cite{deru75} whichwas a major advance beyond the naive quark
model and it is presented in many textbooks. The spin-color or
Fermi-Breit (FB) model,\cite{libe77} although it leads to the same 
massformulas for conventional hadrons and it is still 
phenomenological, incorporates the hypothesis of color. In paper of 
Glozman and Riska\cite{gloz96} it was shown that spin-flavor HFI is 
suitable for light quarks. They analyze a baryon spectrum in terms 
of an SU(3) flavor-symmetric quark-quark interaction that describes 
chiral pseudoscalar boson exchange. In this paper, the hyperfine
interactions (HFIs) between constituent quarks in mesons and 
baryons, including those with one $b$-quark, are used to investigate
how their masses are affected by these interactions and to compare
theoretically obtained masses with experimentally measured ones. The two
types of interactions: Fermi-Breit(FB)
\cite{deru75,libe77,fran81,luch91,silv92,bork08,bork10a,bork12} and 
Glozman-Riska (GR)\cite{gloz95,gloz96,bork07,bork08,bork10a,bork10b} 
(i.e. color-spin and flavor-spin) are used to obtain meson and 
baryon mass formulas. Then we compare our theoretical calculations 
with known masses of $b$-hadrons.

Detailed physical justification of the models is beyond the scope of
this paper. For example, the models are non-relativistic and they
neglect the kinetic energy. Nevertheless, they have several
advantages, e.g. they are rather explicit since they yield 
elementary expressions for the hadron masses, they have few free 
parameters etc. Any more realistic model necessarilly introduces 
additional parameters; even the original Rujula-Georgi-Glashow model 
contained additonal terms with respect to the FB and GR models used 
here (see Ref.~\refcite{deru75} or formula (7.75) in 
Ref.~\refcite{fayy00}). Moreover, bound states in a realistic 
two-body potential can only be found numerically so the dependence of 
theoretical hadron masses on the parameters of the model is not as 
clear as in the simple models. A model with two additional free 
parameters -- given that few (statistical) degrees of freedom would
be left -- would have to give a very good fit to the observed
hadron masses in order to be considered satisfactory.

Although FB and GR HFI are well known,\cite{deru75,gloz95} we 
use these interactions for determining the constituent quark masses 
(especially $b$ quark mass) since we were able to find few papers 
where constituent $b$ quark mass is 
determined.\cite{corc04,chet12,bode12,luch13,peni14,bern14,graj13} 
Also, our approach differs from the ones used in those papers. 
Moreover, the `consituent' quark mass as used here is not exactly the 
same as the `bottom $1S$' mass often defined (using bottomonium) in 
such determinations, e.g. in Ref.~\refcite {hoan12}. We also provide 
explicit mass formulas for hadrons containing $b$ quark using both 
HFIs.

In this paper we make one of the first attempts to estimate
uncertainties of the constituent quark masses. This estimation was
partly motivated by the need to discern between the physically
distinct FB and GR HFI because both had yielded satisfactory fits of
the hadron masses in some cases.\cite{bork10a}

One of our goals is to investigate the models that are simple
enough to analyze the exotic hadron states -- primarily the tetraquarks --
because more elaborate models, having far less transparent 
dependence on model parameters, become quite difficult to analyze 
(if not to compute) in systems with several quark pairs.

This paper is organized as follows: in Sec.~\ref{sec2} we present
the two strong hyperfine interactions, then in Sec.~\ref{sec3}
we give their influence on hadron masses and derive formulas; the
method of our calculation of the constituent quark masses and the
coupling constants (least-square fit) is given in Sec.~\ref{sec4},
in Sec.~\ref{sec5} the different combinations of hadron mass
equations are solved (clustering of hadron groups), and in
Sec.~\ref{sec6} we discuss the obtained results. We point out the  
main conclusions in Sec.~\ref{sec7}.

\section{Strong hyperfine interactions and the schematic model}
\label{sec2}

The main interaction which binds quarks into groups (hadrons)
depends on the color and the spin. With no hyperfine interaction
added, there would be degenerate hadrons with different spins. To
avoid this spin degeneration, hyperfine interaction is included and
it depends, among other properties, on the spin too.

Strong Fermi-Breit hyperfine interaction Hamiltonian\cite{libe77}
with SU(3) flavor symmetry breaking is of the form:
\begin{equation}
H_\mathrm{FB} = C \sum\limits_{i < j}
\left( {\dfrac{\vec \sigma_i \vec \sigma_j}{m_i m_j}} \right)
\left( {\lambda_i^C \lambda_j^C} \right),
\label{equ01}
\end{equation}

\noindent where $m_i$ are constituent masses of the interacting
quarks, $\sigma_i$ are the Pauli spin matrices, $\lambda^C_i$ are
the color Gell-Mann matrices and $C$ is a constant. This interaction
is also called color-spin interaction. As explained in the
papers\cite{deru75,schn75}, Fermi-Breit interaction originates from
one gluon exchange between two bodies, in analogy with the photon
exchange between charged Dirac particles. The Fermi term of this
interaction refers to hyperfine splitting of masses, i.e. it depends 
on inverse product of the quark masses, while Breit interaction 
contains a part which is spin-dependent (short range gluon 
interactions) and another spin-independent part (forces that keep $q 
\bar{q}$ pairs in color singlets). In the case of this interaction, 
we neglect all other potentials in the system, and include only 
Fermi-Breit two particle interaction.

The formulas derived from (\ref{equ01}) reduce to the simplified 
Rujula-Georgi-Glashow model as the expectation values of the products 
$\lambda_i^C \lambda_j^C$ can be absorbed into the HFI constants. 
However, the models are not equivalent; not only does $H_\mathrm{FB}$ 
include the hypothesis of color explicitly, but it also leads to 
different mass formulas for exotic hadrons since the products 
$\lambda_i^C \lambda_j^C$ are different for $q q$ and $q\bar{q}$ 
pairs. 

Strong Glozman-Riska Hamiltonian\cite{gloz96} is of the form:
\begin{equation}
H_\mathrm{GR} = - C_\chi \sum\limits_{i < j}
\left( -1 \right)^{\mathop \alpha_{ij}}
\left( {\dfrac{\vec \sigma_i \vec \sigma_j}{m_i m_j}}\right)
\left( {\lambda_i^F \lambda_j^F } \right); \ \ \ 
\left( -1 \right)^{\mathop \alpha_{ij}} = \left\{
{\begin{array}{*{20}c} {\begin{array} {*{20}c}{ -1,} & {q\bar q} \\
\end{array}} \\
{\begin{array}{*{20}c} { +1,} & {qq \ {\rm or}\ \bar q\bar q} \\
\end{array}} \\
\end{array}} \right\},
\label{equ02}
\end{equation}

\noindent where $\lambda_{i}^{F}$ are Gell-Mann matrices for flavor
SU(3), $\sigma_{i}$ are the Pauli spin matrices and C$_{\chi}$ is a
constant. This is flavor-spin interaction. This interaction between
constituent quarks describes pseudoscalar boson exchange, i.e. fine
structure of the spectrum is based on the interaction mediated by
the SU(3)$_\mathrm{F}$ octet of pseudoscalar mesons, which are the
Goldstone bosons.

We employ these schematic color-spin and flavor-spin interactions
between quarks and antiquarks which lead to hyperfine interaction
contributions to the meson and baryon masses. The schematic
approximation means that we used two-particle interaction: in our
calculation of hadron masses, we pay attention only to the
short-range forces which arise from one-gluon exchange, i.e. the
hadron masses are described in terms of two-body quark-quark forces.

\section{Hadron masses with FB and GR HFI\lowercase{s}}
\label{sec3}

The contribution of HFI to hadron masses would be
$m_{\nu,\mathrm{HFI}} = \left\langle \nu \right|\left\langle \chi
\right|H_\mathrm{HFI} \left| \chi \right\rangle \left| \nu
\right\rangle$, where $\chi$ denotes the spin wave function and
$\nu$ the flavor wave function while HFI is either FB or GR
interaction. For total hadron masses $m_{\nu}$ we have $m_{\nu} =
m_{\nu,0} + m_\mathrm{\nu,HFI}$, where $m_{\nu,0}$ are masses
without influence of HFI.

Experimentally detected hadrons are listed in the Summary Tables of
the Particle Data Group (PDG).\cite{PDG12} Among them, we choose
the particles with orbital momentum $L$ = 0, and with a certain
total momentum $J = L + S$ ($S$ being spin) and the parity $P$ (note
that $P = (-1)^{L+1}$ for mesons and $P = (-1)^{L}$ for baryons).

Among the mesons listed in Particle Physics Summary Tables, we choose
the following particles:

\begin{itemize}

\item light pseudoscalar mesons $J^P = 0^-$; $S = 0$:

$\pi^+$, $\pi^0$, $\pi^-$, $K^+$, $K^0$, $\bar{K}^0$, $K^-$
(note that we did not take into account $\eta$ and $\eta'$
because their mixing changes their properties as well as their
masses),
\item light vector mesons $J^P = 1^-$; $S = 1$:

$\rho^+$, $\rho^0$, $\rho^-$, $K^{*+}$, $K^{*0}$, $\bar{K}^{*0}$, 
$K^{*-}$, $\omega$, $\phi$,

\item bottom mesons $J^P = 0^-$; $J^P = 1^-$:

$B^{+}$, $B^{0}$, $\bar{B}^{0}$, $B^{-}$, $B^{*}$,

\item strange bottom mesons $J^P = 0^-$; $J^P = 1^-$:

$B_S^0$, $\bar{B}_S^0$, $B_S^*$.
\end{itemize}

We first present equations for theoretical meson masses with FB HFI
included obtained from
(\ref{equ01}). We denote the constant for this interaction for
mesons by $C^m$. The possible small mass difference between $u$ and
$d$ constituent quarks is neglected. However, the difference in the 
observed masses within the isospin multiplets is taken into account 
(except for $\Delta$-baryons, for which an average mass is used) so, 
in those cases, the same theoretical masses are fitted to different 
observed ones.

\begin{equation}
m^{\rm th}_{\pi^{\pm}} = 2m_u - \dfrac{3C^{m}}{m_u^2} = m^{\rm th}_{\pi^0} ,\qquad
 m^{\rm th}_{K^{\pm}} = m_u + m_s - \dfrac{3C^{m}}{m_u m_s} = m^{\rm th}_{K^0} = m^{\rm th}_{\bar{K}^0}.
\label{equ03}
\end{equation}

\begin{equation}
\begin{array}{l}
m^{\rm th}_{\rho^{\pm}} = m^{\rm th}_{\rho^0} = 2m_u + 
\dfrac{C^{m}}{m_u^2} = m^{\rm th}_\omega,\qquad
m^{\rm th}_\phi = 2m_s + \dfrac{C^{m}}{m_s^2}, \\
m^{\rm th}_{K^{*+}} = m^{\rm th}_{K^{*-}} = m_u + m_s + \dfrac{C^{m}}{m_u m_s} = m^{\rm th}_{K^{*0}} = m^{\rm th}_{\bar{K}^{*0}}.
\end{array}
\label{equ04}
\end{equation}

\begin{equation}
m^{\rm th}_{B^+} = m^{\rm th}_{\bar{B}^-}= m_u + m_b - \dfrac{3C^{m}}{m_u m_b} = m^{\rm th}_{B^0} = m^{\rm th}_{\bar{B}^0}, \qquad
 m^{\rm th}_{B^*} = m_u + m_b + \dfrac{C^{m}}{m_u m_b}.\!
\label{equ05}
\end{equation}

\begin{equation}
m^{\rm th}_{B_s^0} = m_s + m_b - \dfrac{3C^{m}}{m_s m_b} = m^{\rm th}_{\bar{B}_s^0}, \qquad
m^{\rm th}_{B_s^*} = m_s + m_b + \dfrac{C^{m}}{m_s m_b}.
\label{equ06}
\end{equation}

Now we give masses with GR from (\ref{equ02}) where the constant is 
denoted by $C_{\chi}^m$.

\begin{equation}
m^{\rm th}_{\pi^{\pm}} = 2m_u - \dfrac{2C_{\chi}^{m}}{m_u^2} = m^{\rm th}_{\pi^0}, \qquad
 m^{\rm th}_{K^{\pm}} = m_u + m_s - \dfrac{2C_{\chi}^{m}}{m_u m_s}
 = m^{\rm th}_{K^0} = m^{\rm th}_{\bar{K}^0}.
\label{equ07}
\end{equation}

\begin{equation}
\begin{array}{l}
m^{\rm th}_{\rho^{\pm}} = m^{\rm th}_{\rho^0} = 2m_u + 
\dfrac{2C_{\chi}^{m}}{3m_u^2} = m^{\rm th}_\omega,\qquad
m^{\rm th}_\phi = 2m_s - \dfrac{16C_{\chi}^{m}}{3m_s^2}, \\
m^{\rm th}_{K^{*+}} = m^{\rm th}_{K^{*-}} = m_u + m_s + \dfrac{2C_{\chi}^{m}}{3m_u m_s} = m^{\rm th}_{K^{*0}} = m^{\rm th}_{\bar{K}^{*0}}.
\end{array}
\label{equ08}
\end{equation}

\begin{equation}
m^{\rm th}_{B^+} = m^{\rm th}_{\bar{B}^-}= m_u + m_b - \dfrac{2C_{\chi}^{m}}{m_u m_b} = m^{\rm th}_{B^0} = m^{\rm th}_{\bar{B}^0}, \qquad
m^{\rm th}_{B^*} = m_u + m_b + \dfrac{2C_{\chi}^{m}}{3 m_u m_b}.\!
\label{equ09}
\end{equation}

\begin{equation}
m^{\rm th}_{B_s^0} = m_s + m_b - \dfrac{2C_{\chi}^{m}}{m_s m_b} = m^{\rm th}_{\bar{B}_s^0}, \qquad
m^{\rm th}_{B_s^*} = m_s + m_b + \dfrac{2C_{\chi}^{m}}{3m_s m_b}.
\label{equ10}
\end{equation}

From the baryons listed by the PDG,\cite{PDG12} we choose these 
particles (note that the subscripts indicate heavy quark content, 
and in our case subscript $b$ indicates content of one $b$ quark):
\begin{itemize}
\item light baryons - octet with mixed symmetry $J^P = 1/2^+$:

$p$, $n$, $\Sigma^+$, $\Sigma^0$, $\Sigma^-$, $\Xi^0$, $\Xi^-$, 
$\Lambda$,

\item light baryons - symmetric decuplet $J^P = 3/2^+$:

$\Delta^{++}$, $\Delta^+$, $\Delta^0$, $\Delta^-$, $\Sigma^{*+}$,
$\Sigma^{*0}$, $\Sigma^{*-}$, $\Xi^{*0}$, $\Xi^{*-}$, $\Omega$,

\item bottom baryons $J^P = 1/2^+$; $J^P = 3/2^+$:

$\Sigma_b^+$, $\Sigma_b^-$, $\Lambda_b$, $\Sigma_b^{*+}$,
$\Sigma_b^{*-}$, $\Omega_b$.
\end{itemize}

We give their theoretical masses with FB HFI influence by
Eqs.~(\ref{equ11})-(\ref{equ13}). Constant for the FB HFI for
baryons is denoted by $C^b$.

\begin{equation}
\begin{array}{l}
m^{\rm th}_p = 3m_u - 3C^b \dfrac{1}{2m_u ^2} = m^{\rm th}_n,\qquad
m^{\rm th}_\Lambda = 2m_u + m_s - 3C^b \dfrac{1}{2m_u^2},\\
m^{\rm th}_{\Sigma^+} = 2m_u + m_s + 2C^b \dfrac{1}{m_u^2}\left( \dfrac{1}{4}- \dfrac{m_u}{m_s} \right) =
m^{\rm th}_{\Sigma^0} = m^{\rm th}_{\Sigma^-}, \\
m^{\rm th}_{\Xi^0} = m_u + 2m_s + 2C^b \dfrac{1}{m_s^2}\left( \dfrac{1}{4} -\dfrac{m_s}{m_u} \right) = m^{\rm th}_{\Xi^-}.
\end{array}
\label{equ11}
\end{equation}

\begin{equation}
\begin{array}{l}
m^{\rm th}_{\Delta^{++}} = m^{\rm th}_{\Delta^+} = m^{\rm th}_{\Delta^0} = m^{\rm th}_{\Delta^-} = 3m_u + 3C^b \dfrac{1}{2m_u^2},\\
m^{\rm th}_{\Sigma^{*+}} = 2m_u + m_s + C^b \dfrac{1}{m_u^2} \left(
\dfrac{1}{2} + \dfrac{m_u}{m_s} \right) =
m^{\rm th}_{\Sigma^{*0}} = m^{\rm th}_{\Sigma^{*-}}, \\
m^{\rm th}_{\Xi^{*0}} = m_u + 2m_s + C^b \dfrac{1}{m_s^2} \left( \dfrac{1}{2} + \dfrac{m_s}{m_u} \right) = m^{\rm th}_{\Xi^{*-}},\\
m^{\rm th}_\Omega = 3m_s + 3C^b \dfrac{1}{2m_s^2}.
\end{array}
\label{equ12}
\end{equation}

\begin{equation}
\begin{array}{l}
m^{\rm th}_{\Sigma^+_{b}} = 2m_u + m_b + 2C^b \dfrac{1}{m_u^2} \left(
\dfrac{1}{4} - \dfrac{m_u}{m_b} \right) = m^{\rm th}_{\Sigma^-_{b}},\\
m^{\rm th}_{\Lambda_b} = 2m_u + m_b - 3C^b \dfrac{1}{2m_u^2}, \\
m^{\rm th}_{\Sigma^{*+}_{b}} = 2m_u + m_b + C^b \dfrac{1}{m_u ^2} \left(
\dfrac{1}{2} + \dfrac{m_u}{m_b} \right) = m^{\rm th}_{\Sigma^{*-}_{b}},\\
m^{\rm th}_{\Omega_b} = 2m_s  + m_b + C^b \dfrac{1}{m_s^2} \left(
\dfrac{1}{2} + \dfrac{m_s}{m_b} \right).
\end{array}
\label{equ13}
\end{equation}

Now we give masses with GR HFI. The GR HFI constant is denoted by 
$C_{\chi}^b$.

\begin{equation}
\begin{array}{l}
m^{\rm th}_p = 3m_u - 8C_{\chi}^{b} \dfrac{1}{m_u ^2} = m^{\rm th}_n, \quad m^{\rm th}_\Lambda = 2m_u + m_s - C_{\chi}^{b} \dfrac{1}{3m_u^2}\left( 13 +\dfrac{11m_u}{m_s} \right),\\
m^{\rm th}_{\Sigma^+} = 2m_u + m_s - C_{\chi}^{b} \dfrac{1}{m_u^2}\left( 1
+ \dfrac{7m_u}{m_s} \right) =
m^{\rm th}_{\Sigma^0} = m^{\rm th}_{\Sigma^-}, \\
m^{\rm th}_{\Xi^0} = m_u + 2m_s - C_{\chi}^{b} \dfrac{1}{m_s^2}\left( 1 +
\dfrac{7m_s}{m_u} \right) = m^{\rm th}_{\Xi^-}.
\end{array}
\label{equ14}
\end{equation}

\begin{equation}
\begin{array}{l}
m^{\rm th}_{\Delta^{++}} = m^{\rm th}_{\Delta^+} = m^{\rm th}_{\Delta^0} = m^{\rm th}_{\Delta^-} = 3m_u - \dfrac{4C_{\chi}^{b}} {m_u^2}, \\
m^{\rm th}_{\Sigma^{*+}} = 2m_u + m_s - 8C_{\chi}^{b} \dfrac{1}{3m_u^2} \left(\dfrac{1}{2} + \dfrac{m_u}{m_s} \right) =
m^{\rm th}_{\Sigma^{*0}} = m^{\rm th}_{\Sigma^{*-}}, \\
m^{\rm th}_{\Xi^{*0}} = m_u + 2m_s - 8C_{\chi}^{b} \dfrac{1}{3m_s^2} \left(
\dfrac{1}{2} + \dfrac{m_s}{m_u} \right) = m^{\rm th}_{\Xi^{*-}}, \\
m^{\rm th}_\Omega = 3m_s - 4C_{\chi}^{b} \dfrac{1}{m_s^2}.
\end{array}
\label{equ15}
\end{equation}

\begin{equation}
\begin{array}{l}
m^{\rm th}_{\Sigma^+_{b}} = 2m_u + m_b - C_{\chi}^{b} \dfrac{1}{m_u^2}
\left( 1 + \dfrac{7m_u}{m_b} \right) = m^{\rm th}_{\Sigma^-_{b}}, \\
m^{\rm th}_{\Lambda_b} = 2m_u + m_b - C_{\chi}^{b} \dfrac{1}{3m_u^2} \left(
13 + \dfrac{11m_u}{m_b} \right), \\
m^{\rm th}_{\Sigma^{*+}_{b}} = 2m_u + m_b - 8C_{\chi}^{b} \dfrac{1}{3m_u^2}
\left( \dfrac{1}{2} + \dfrac{m_u}{m_b} \right) = m^{\rm th}_{\Sigma^{*-}_{b}}, \\
m^{\rm th}_{\Omega_b} = 2m_s + m_b - 8C_{\chi}^{b} \dfrac{1}{3m_s^2} \left(
\dfrac{1}{2} + \dfrac{m_s}{m_b} \right).
\end{array}
\label{equ16}
\end{equation}

\begin{table}[ht!]
\tbl{Fitted values of constituent quark masses $m_u$ (= $m_d$),
$m_s$, $m_b$ (MeV) and the hyperfine constants $C^m$ and $C_\chi^m$
(10$^7$ MeV$^3$), obtained by $\chi^2$ fits of meson masses with FB and GR HFI.}
{\begin{tabular}{llccccc}
\hline\noalign{\smallskip} Fit & HFI & Mesons &
\multicolumn{3}{c}{Quark masses (MeV)} &
Constant \\
No. & & & $m_u$ = $m_d$ & $m_s$ & $m_b$ & ($\times 10^7$
$\mathrm{MeV^3}$) \\
\hline
& & & & & \\
1 & FB & $\pi$, $K$, $\rho$, $K^*$, $\omega$, $\phi$,  & 307.54 $\pm$ 1.16 & 487.41 $\pm$ 1.56
& 4967.20 $\pm$ 18.73 & 1.50 $\pm$ 0.02 \\
& GR & $B$, $B^*$,
$B_S$, $B_S^*$ & 293.09 $\pm$ 11.61 & 513.00 $\pm$ 15.84 & 4964.87 $\pm$
189.22 & 1.93 $\pm$ 0.26 \\
\noalign{\smallskip}
2 & FB & $\pi$, $K$, $\rho$, $K^*$, $\omega$, $\phi$ & 307.49 $\pm$
1.19 & 487.52 $\pm$ 1.59 & - & 1.50 $\pm$ 0.02 \\
& GR & & 293.05 $\pm$ 14.36 & 513.15 $\pm$ 19.59 & - & 1.93 $\pm$
0.32 \\
\noalign{\smallskip}\hline
\end{tabular}}
\label{tab01}
\end{table}

\begin{table}[ht!]
\tbl{The same as Table \ref{tab01}, but for baryon fit.}
{\begin{tabular}{llccccc} \hline\noalign{\smallskip} Fit & HFI &
Baryons & \multicolumn{3}{c}{Quark masses (MeV)} &
Constant \\
No. & & & $m_u$ = $m_d$ & $m_s$ & $m_b$ & ($\times 10^7$
$\mathrm{MeV^3}$) \\
\hline
& & & & & \\
1 & FB & $p$, $n$, $\Sigma$, $\Xi$, $\Lambda$, $\Delta$, $\Sigma^*$,
& 363.03 $\pm$ 0.87 & 538.71 $\pm$ 1.69 & 5043.15 $\pm$ 15.08 & 1.30
$\pm$ 0.03 \\
& GR & $\Xi^*$, $\Omega$, $\Sigma_b$, $\Lambda_b$, $\Sigma_b^*$, $\Omega_b$ 
& 500.11 $\pm$ 3.44 & 624.39 $\pm$ 3.33 & 4923.73 $\pm$ 24.25
& 1.72 $\pm$ 0.07 \\
\noalign{\smallskip}
2 & FB & $p$, $n$, $\Sigma$, $\Xi$, $\Lambda$, $\Delta$, $\Sigma^*$,
 & 362.94 $\pm$ 0.94 & 538.89 $\pm$ 1.85 & - & 1.30
$\pm$ 0.03 \\
& GR & $\Xi^*$, $\Omega$ & 499.97 $\pm$ 3.89 & 624.32 $\pm$ 3.78 & - & 1.71 $\pm$ 0.07
\\
\noalign{\smallskip}\hline
\end{tabular}}
\label{tab02}
\end{table}

\section{Fit of constituent quark masses}
\label{sec4}

For calculating constituent quark masses we used theoretical 
equations for meson and baryon masses given in Sec.~\ref{sec3}. We 
derived theoretical formulas for FB and GR HFI contribution and then 
obtained the total theoretical masses of the hadrons we have
studied. The corresponding experimental values of the hadron masses
are taken from PDG data.\cite{PDG12}

For our calculations we used multidimensional least-square fit of
quark masses and hyperfine constant, using subroutine ''lfit'' from
Numerical Recipes in FORTRAN,\cite{pres92} modified according to
the instructions in the last paragraph of Sec.~15.4 of 
Ref.~\refcite{pres92}. The equations for hadron masses are 
linearized by expansion in Taylor series (up to the first order) so 
we obtained the system of linear equations for differences between 
experimental and theoretical hadron masses. The unknown variables in 
these equations are corrections to the parameters (i.e. corrections 
to the constituent quark masses and the constant of hyperfine 
interaction), which are obtained by linear least-square fitting. The 
uncertainties are estimated during the fitting procedure as square 
roots of the corresponding diagonal elements of covariance matrix, 
according to Eq. (15.4.15) of Ref.~\refcite{pres92}.

The fitting is performed by minimizing $\chi^2$ between the
theoretical and experimental hadron masses, according to the
following procedure:
\begin{itemize}
\item [-] First, initial values for the constituent quark
    masses are assumed and used to calculate the initial values
    of the theoretical masses of hadrons (mesons and baryons),
    and $\chi^2$ between theoretical and experimental hadron
    masses;

\item [-] In the next iteration the corrections to the
    constituent quark masses and the constant of hyperfine
    interaction are obtained by least-square fitting, and used
    to obtain new values of the parameters by adding these
    corrections to the estimates from the previous iteration.
    The new values of the theoretical hadron masses are then
    obtained from these corrected parameters, as well as a new
    value of $\chi^2$ between theoretical and experimental
    hadron masses. We repeat this procedure until the fit
    converges, i.e. while $\chi^2$ decreases, and finally,

\item [-] we choose the set of the constituent quark masses and
    the constant which gives the least $\chi^2$.
\end{itemize}

Assuming that the number of fitted hadrons (mass equations) is $N$,
and the number of unknown parameters (constituent quark masses and
the constant) is $m$, we used the following expression for reduced
$\chi^2$:
\begin{equation}
\chi^2 = \dfrac{1}{N-m}\sum \limits_{i = 1} \limits^N
{\dfrac{y_i^2}{\sigma_i^2}},
\label{equ17}
\end{equation}

\noindent with $y_i$ being the $i$-th difference between the
measured and theoretical hadron masses, and $\sigma_i$ the $i$-th
standard deviation:
\begin{equation}
\sigma_i^2=\sigma_{i,\mathrm{exp}}^2+\sigma_{i,\mathrm{theor}}^2,
\label{equ18}
\end{equation}

\noindent where $\sigma_{i,\mathrm{exp}}$ is an experimental
standard deviation given by PDG,\cite{PDG12} and
$\sigma_{i,\mathrm{theor}}$ is a theoretical standard deviation. We
added $\sigma_{i,\mathrm{theor}}$ in quadrature with the
experimental errors to avoid having the fit to experiment
arbitrarily dominated by the most accurate measurements 
(Ref.~\refcite{franpc}, see also Ref.~\refcite{fran02}). We took
$\sigma_{i,\mathrm{theor}}$ to be proportional to the experimental
masses ($\sigma_{i,\mathrm{theor}}=A \cdot m_{exp}$), and chose the
constant of proportionality $A$ in such a way to yield the reduced
$\chi^2$ as close as possible to 1.

In brief, we fitted four parameters: $m_u$, $m_s$, $m_b$, $C$, so
that the $\chi^2$ between measured and theoretical masses is
minimized. As we mentioned in Ref. \refcite{bork10a}, for every
system of equations the fast convergence is achieved, even in the
case when initial values of parameters differ much from their final
values, which tells about goodness of our theoretical model and the
fitting method.

In Tables \ref{tab01} and \ref{tab02} we give the results from 
meson and baryon fits, when HFIs are included. The calculation is 
done by combining the hadrons we chose in Sec.~\ref{sec4}. In 
equations for meson masses there is hyperfine constant labeled by 
$C^m$, and for baryon masses there is hyperfine constant labeled by 
$C^b$. These two constants are not equal, but we can say that they 
are of the same order of magnitude, $C^m \sim C^b$.

\begin{table}[ht!]
\tbl{Absolute differences, in MeV, between experimental masses of
mesons and our calculated theoretical masses with FB HFI (third
column) and GR HFI (fourth column).
Parameter values used here are obtained by $\chi^2$ fit of all
mesons (see fit No. 1 in Table \ref{tab01} for the third and fourth
column).}
{\begin{tabular}{lrrr} \hline\noalign{\smallskip} Meson &
$m$ (MeV) & $\Delta m_\mathrm{FB}$ & $\Delta
m_\mathrm{GR}$ \\
\hline
\hline\noalign{\smallskip}
$\pi^{\pm}$ & 139.57 $\pm$ 0.01 & 1.51 & 3.29 \\
$\pi^0$ & 134.98 $\pm$ 0.01 & 3.09 & 1.30  \\
$K^{\pm}$ & 493.67 $\pm$ 0.02 & 0.29 & 55.37 \\
$K^0$, $\bar{K}^0$ & 497.61 $\pm$ 0.03 & 3.65 & 51.44 \\
$\rho^{\pm}$, $\rho^0$ & 775.26 $\pm$ 0.25 & 1.18 & 39.11 \\
$K^{*\pm}$ & 891.66 $\pm$ 0.26 & 3.61 & 0.11 \\
$K^{*0}$, $\bar{K}^{*0}$ & 895.81 $\pm$ 0.19 & 0.54 & 4.04 \\
$\omega$ & 782.65 $\pm$ 0.12 & 8.57 & 46.50 \\
$\phi$ & 1019.46 $\pm$ 0.02 & 18.66 & 385.06 \\
$B^{\pm}$ & 5279.25 $\pm$ 0.17 & 34.04 & 47.85 \\
$B^0$, $\bar{B}^0$ & 5279.58 $\pm$ 0.17 & 34.37 & 48.18 \\
$B^*$ & 5325.20 $\pm$ 0.40 & 40.62 & 58.38 \\
$B_S^0$, $\bar{B}_S^0$  & 5366.77 $\pm$ 0.24 & 69.20 & 95.93 \\
$B_S^*$ & 5415.40 $\pm$ 2.20 & 45.42 & 67.53 \\
\noalign{\smallskip}\hline
\end{tabular}}
\label{tab03}
\end{table}

\begin{table}[ht!]
\tbl{The same as Table \ref{tab03}, but for baryons (see fit No. 1
in Table \ref{tab02} for the third and fourth column).}
{\begin{tabular}{lrrr} \hline\noalign{\smallskip} Baryon & $m$ (MeV)
& $\Delta m_\mathrm{FB}$ & $\Delta
m_\mathrm{GR}$ \\
\hline
$p$ & 938.27 $\pm$ 0.01 & 3.28 & 13.15 \\
$n$ & 939.57 $\pm$ 0.01 & 1.99 & 11.86 \\
$\Sigma^+$ & 1189.37 $\pm$ 0.07 & 7.98 & 18.07 \\
$\Sigma^0$ & 1192.64 $\pm$ 0.03 & 11.25 & 21.34 \\
$\Sigma^-$ & 1197.45 $\pm$ 0.03 & 16.06 & 26.15 \\
$\Xi^0$ & 1314.86 $\pm$ 0.20 & 15.38 & 5.32 \\
$\Xi^-$ & 1321.71 $\pm$ 0.07 & 8.53 & 1.53 \\
$\Lambda$ & 1115.68 $\pm$ 0.01 & 1.56 & 10.10 \\
$\Delta$ (mean) & 1232.00 $\pm$ 4.00 & 4.61 & 6.12 \\
$\Sigma^{*+}$ & 1382.80 $\pm$ 0.35 & 2.58 & 3.78 \\
$\Sigma^{*0}$ & 1383.70 $\pm$ 1.00 & 3.48 & 2.88 \\
$\Sigma^{*-}$ & 1387.20 $\pm$ 0.50 & 6.98 & 0.62 \\
$\Xi^{*0}$ & 1531.80 $\pm$ 0.32 & 2.74 & 11.85 \\
$\Xi^{*-}$ & 1535.00 $\pm$ 0.60 & 5.94 & 8.65 \\
$\Omega$ & 1672.45 $\pm$ 0.29 & 10.69 & 24.65 \\
$\Sigma_b^+$ & 5811.30 $\pm$ 1.90 & 7.09 & 4.74 \\
$\Sigma_b^-$ & 5815.50 $\pm$ 1.80 & 11.29 & 8.94 \\
$\Lambda_b$ & 5619.40 $\pm$ 0.60 & 2.28 & 32.79 \\
$\Sigma_b^{*+}$ & 5832.10 $\pm$ 1.90 & 6.65 & 18.21 \\
$\Sigma_b^{*-}$ & 5835.10 $\pm$ 1.90 & 9.65 & 21.21 \\
$\Omega_b$ & 6071.00 $\pm$ 40.0 & 76.68 & 27.94 \\
\noalign{\smallskip}\hline
\end{tabular}}
\label{tab04}
\end{table}

To compare these two interactions, we calculated the differences
between experimental and theoretical masses of the fitted hadrons,
and their absolute values presented in Tables \ref{tab03} 
and \ref{tab04}. The CPT theorem was not assumed, e.g. $\pi^+$ and 
$\pi^-$ were considered as different points in the meson fits. 

\section{Clustering of hadron groups}
\label{sec5}

\begin{table}[ht!]
\tbl{Calculated constituent quark masses $m_u$ (= $m_d$), $m_s$,
$m_b$ (MeV) and the hyperfine constants $C^m$ and $C_\chi^m$ (10$^7$
MeV$^3$) obtained from clustering of equations for meson masses with
FB (upper rows) and GR (lower rows) HFI.}
{\begin{tabular}{llccccc}
\hline\noalign{\smallskip} System & HFI & Combinations &
\multicolumn{3}{c}{Quark masses
(MeV)} & Constant \\
No. & & of mesons & $m_u$ = $m_d$ & $m_s$ & $m_b$ & ($\times 10^7$
$\mathrm{MeV^3}$) \\
\hline
& & & & & \\
1 & FB & $\pi^+$, $\rho^+$ & 308.17 & -- & -- & 1.51 \\
& GR & & 308.17 & -- & -- & 2.26 \\
\noalign{\smallskip}
2 & FB & $K^+$, $K^{*+}$, $\omega$ & 316.60 & 475.57 & -- & 1.50 \\
& GR & & 316.60 & 475.57 & -- & 2.25 \\
\noalign{\smallskip}
3 & FB & $B^+$, $B^*$ & -- & -- & 5001.21 & 1.80 \\
& GR & & -- & -- & 5001.21 & 2.70 \\
\noalign{\smallskip}
4 & FB & $\pi^+$, $K^+$, $B_S^0$ & -- & 490.47 & 4896.04 & 1.58 \\
& GR & & -- & 490.47 & 4896.04 & 2.37 \\
\noalign{\smallskip}\hline
\end{tabular}}
\label{tab05}
\end{table}

\begin{table}[ht!]
\tbl{The same as Table \ref{tab05}, but for baryons.}
{\begin{tabular}{llccccc} \hline\noalign{\smallskip} System & HFI &
Combinations & \multicolumn{3}{c}{Quark masses (MeV)}
& Constant \\
No. & & of baryons & $m_u$ = $m_d$ & $m_s$ & $m_b$ & ($\times 10^7$
$\mathrm{MeV^3}$) \\
\hline
& & & & & \\
1 & FB & $p$, $\Delta^{++}$ & 361.71 & -- & -- & 1.28 \\
& GR & & 508.58 & -- & -- & 1.90 \\
\noalign{\smallskip}
2 & FB & $\Sigma^+$, $\Xi^0$, $\Sigma^{*+}$ & 375.41 & 522.62 & -- &
1.27 \\
& GR & & 495.85 & 602.10 & -- & 1.47 \\
\noalign{\smallskip}
3 & FB & $\Sigma_b^+$, $\Lambda_b$ & -- & -- & 5036.70 & 1.41 \\
& GR & & -- & -- & 4917.23 & 1.62 \\
\noalign{\smallskip}
4 & FB & $\Sigma_b^+$, $\sigma_b^{*+}$, $\omega_b$ & -- & 500.15 &
5039.82 & 1.29 \\
& GR & & -- & -- & -- & -- \\
\noalign{\smallskip}\hline
\end{tabular}}
\label{tab06}
\end{table}

\begin{figure}
\centering
\includegraphics[width=0.98\textwidth]{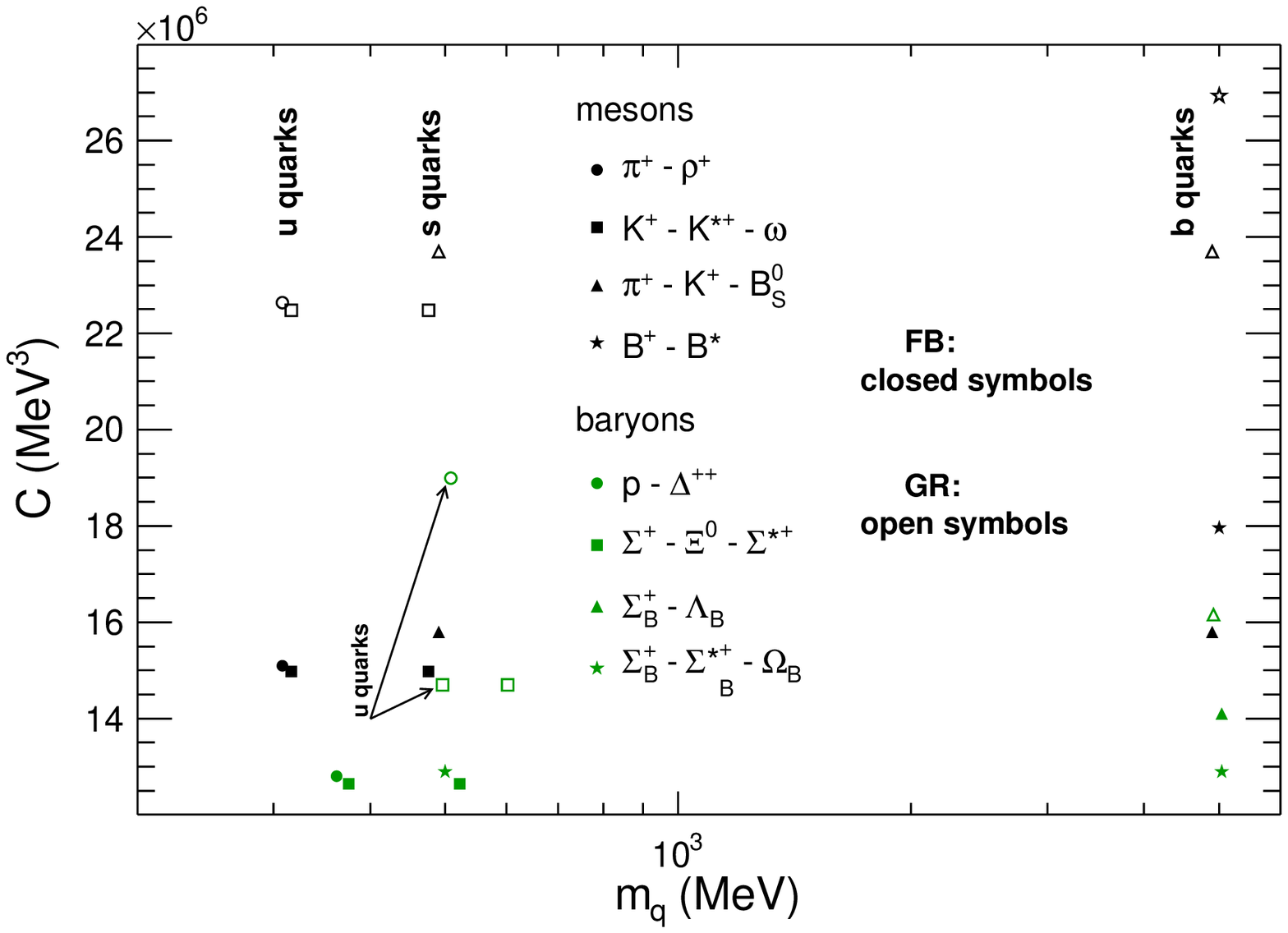}
\caption{Masses of constituent quarks confined in different hadrons
and the related constants obtained as analytical solutions of the
systems of two and three equations formed from sets of
Eqs.~(\ref{equ03}) to (\ref{equ16}). The results for mesons
(baryons) are depicted with closed black (green) symbols for FB HFI,
while in the case of GR HFI they are presented with corresponding
open symbols.}
\label{fig01}
\end{figure}

Along with the least-square fit method described above, another
approach has been applied too. Based on similarity of quark content,
the whole set of equations, used in the $\chi^2$-fitting method, has
been divided into a certain number of subsets chosen in such a way
to form a minimal system of equations which could be analytically 
solved. The Mathematica 9.0 software has been used to solve these 
systems of equations for mesons and baryons. Typically, there were 
sets of two equations with two unknown parameters and the sets of 
three equations with three unknown parameters. For example, from set 
of Eqs. (\ref{equ03}) and (\ref{equ04}) a system of equations for
$\pi^+$ and $\rho^+$ has been formed. This is then analytically
solved in $m_u$ and $C^m$. In the case when three equations are
chosen to form the system of equations, like for example equations
for $\Sigma^{+}$, $\Xi^{0}$ and $\Sigma^{*+}$ taken from
Eqs.~(\ref{equ11}) and (\ref{equ12}), analytically one can get
solutions in $m_u$, $m_s$ and $C^b$. In this analysis, four systems
of equations for mesons and four for baryons have been formed in the
case of FB HFI. The same is done in the case of GR HFI, except for
baryons where system of equations formed for $\Sigma_{b}^{+}$,
$\Sigma_{b}^{*+}$ and $\Omega_{b}$ did not give real solution. Here,
one should also notice that in the case of hadrons which contain $b$
quark, in order to solve equation it was necessary to include
numerical value of $m_{u}$ previously obtained from systems of light
hadrons. That value is obtained as a mean value of $m_{u}$
calculated separately for mesons and for baryons and for both FB
and GR HFI.

As the result, the method described above forms clusters of $m_{q} -
C$ points as can be seen in Fig.~\ref{fig01}. This clustering shows
the way how the constituent quark masses and the strength of the
interaction constants appear in different environments: mesons -
baryons, light hadrons - heavy hadrons within the two analyzed
hyperfine interactions. The biggest formed clusters are those
characterized by approximately the same quark mass for each of
three different quark families ($u$, $b$ and $s$) but with different
constants within the given type of interaction. Here, one could note
that points from both types of interactions belong to the same
cluster for a given quark family. The only exceptions seen are green
open circle and green open square extracted from baryonic equations
for $p$ and $\Delta^{++}$ and from equations for $\Sigma^{+}$,
$\Xi^{0}$ and $\Sigma^{*+}$ respectively which gives masses of
$u$-quarks derived from GR HFI. The corresponding masses are shifted
to the position which belongs to the '$s$-quarks' cluster. Within
these big clusters, one can see smaller clusters which contain
points obtained from mesons or from baryons. For the light quarks,
typically, mesonic clusters have smaller quark masses and bigger
interaction constants with respect to baryonic clusters. Concerning
the masses, this observation is clearly seen for the $u$-quarks,
gets smaller for the $s$-quarks and nearly invisible for the
$b$-quark. On the other hand, the difference between the extracted 
interaction constant in mesonic and baryonic clusters gets slightly 
larger going from light to heavy quarks. A similar clustering can 
also be seen in the case of GR HFI.

The different values of the extracted masses and interaction
constants reflect the different enviroment -- mesonic or
baryonic -- and the type of interaction.

\section{Discussion}
\label{sec6}

The experimental observation is that there are no free quarks but
that they only exist bound in hadrons. This phenomenon is known as
confinement. The constituent quark mass is the effective mass of a
quark, which is only defined if the quark is confined and bound in
the hadron. We calculate constituent quark masses using the improved
fitting procedure\cite{franpc} of groups of mesons and baryons and
using method of clustering of some hadron groups.

We give mass formulas for hadrons containing $b$ quark using FB and
GR HFI. We also calculate coupling constants of these interactions
and showed that they are not equal but are of the same order of
magnitude. We investigate how constituent quark masses, and coupling
constants, depend on different hadron environment and how effective
these two interactions are.

The obtained results show that quark masses depend on the particular
hadron model, and are different for two studied HFIs. On average, FB
interaction gives much better fit: the uncertainties of the constituent
quark masses are greater by an order of magnitude for GR HFI than for 
FB HFI in the case of mesons and by a factor of two in the case of 
baryons. We have to stress that FB interaction is working well for 
heavy-light mesons and baryons if they contain only one heavy quark, 
while GR HFI in some cases failed for heavy baryons. GR HFI also 
fails for $\phi$ meson.

Just for comparison with the results from least-square fit, we also
calculated clustering for some sets of equations and presented it in
Tables \ref{tab05} and \ref{tab06} and in Fig. \ref{fig01}. Figure
shows masses of constituent quarks confined in different hadrons and
the related constants: $x$-axes represents quark mass (in MeV) and
$y$-axes the hyperfine constant (in MeV$^3$). For light mesons
(systems (1) and (2) in Table \ref{tab05}), as well as for heavy
mesons (systems (3) and (4)) FB and GR give very similar results for
constituent quark masses. For $m_u$ and $m_s$ in heavy meson systems
we obtained larger values than in light mesons. For baryons, we can
notice that in case of light baryons (systems (1) and (2) in Table
\ref{tab06}) FB and GR HFI results differ which is opposite from
case of light mesons: GR interaction gives larger values for $m_u$
and $m_s$ than FB. When comparing heavy and light baryons (systems
(3) and (4)) we have the greater value for $C^b$ in heavy baryons
than in light baryons. In Table \ref{tab06}, we do not have values
for GR HFI for the fourth system, because physically realistic 
values could not be obtained when solving the equations. We can 
conclude that FB HFI is more accurate interaction than GR HFI.

According to the clustering procedure, both interactions FB and GR have
a similar behavior, but in the case of heavy baryons FB HFI is better 
because GR HFI did not give good results in some cases (i.e. system 
(4) in Table \ref{tab06}). From this method, we can conclude that 
constituent quark masses are very sensitive to environment 
of different hadrons, as well as the values of constants. Values of 
the constants are somewhat higher in hadrons which contain 
$b$-quarks.

In Refs. \refcite{ronc95a,ronc95b} it was shown that in the
constituent quark model, the Feynman-Hellmann theorem and 
semi-empirical mass formulas can be applied to give useful 
information about the masses of mesons and baryons. We obtained that
the Feynman-Hellmann theorem is working well in case of mesons and
baryons with FB HFI, as well as with GR HFI.

\section{Conclusions}
\label{sec7}

In order to determine the constituent quark masses we have used two
methods: (1) least-square fits of both light and open-bottom heavy
light hadrons, (2) clustering of hadron groups. We improved fitting
procedure used in method (1).\cite{bork10a,franpc}

In the previous work, least-square fit gave similar results for
FB and GR HFI for light mesons.\cite{bork10a} For heavy light
hadrons studied in this paper, we find that FB HFI gives much
better fits. This could have been expected given the 
SU(3)$_\mathrm{F}$ nature of the GR HFI. More appropriate spin-flavor 
HFI for light heavy hadrons will be the subject of further 
investigation.

The FB HFI gives reasonably good fits for all hadrons that we have 
considered -- including the open bottom ones, especially when one 
takes into account the simplicity of the model. Further improvements 
are likely to be achievable only in much more elaborate, 
multiparameter models.

We have confirmed that constituent quark mass depends on the type of the
hadron where quarks are confined and on the particular hadron model. 
We show that, in general, quark mass has a larger value in baryons 
than in mesons. Also, it depends on particular type of mesons and 
baryons, i.e. it is not the same in different mesons, but it is more 
similar than in baryons.

\section*{Acknowledgments}
We wish to acknowledge the support by the Ministry of Education,
Science and Technological Development of the Republic of Serbia 
through the projects 176003 and 171019, and by the Ministry of
Science and Technology of the Republic of Srpska through the project
contract No.~19/6-020/961-210/12.

\end{document}